\documentclass{webofc}
\usepackage[varg]{txfonts}   
\newcommand{\eqb}{\begin{equation}}
\newcommand{\eqe}{\end{equation}}
\newcommand{\dmb}{\begin{displaymath}}
\newcommand{\dme}{\end{displaymath}}

\newcommand{\eab}{\begin{eqnarray}}
\newcommand{\eae}{\end{eqnarray}}

\newcommand{\be}{\begin{equation}}
\newcommand{\ee}{\end{equation}}

\woctitle{International Conference on New Frontiers in Physics}
\begin{document}
\title{Thermodynamics of SU(2) quantum Yang-Mills
theory and CMB anomalies}
%
%

\author{Ralf Hofmann\inst{1,2}\fnsep\thanks{\email{r.hofmann@thphys.uni-heidelberg.de}}
}

\institute{Institut f\"ur Theoretische Physik\\ 
Universit\"at Heidelberg\\ 
Philosophenweg 16\\ 
69120 Heidelberg, Germany
\and 
Institut f\"ur Photonenforschung und Synchrotronstrahlung\\ 
Karlsruher Institut f\"ur Technologie\\ 
Hermann-von-Helmholtz Platz 1\\ 
Eggenstein-Leopoldshafen, Germany
}

\abstract{A brief review of effective SU(2) Yang-Mills thermodynamics in the deconfining phase is given, including the construction of the thermal 
ground-state estimate in terms of an inert, adjoint scalar field $\phi$, 
based on non-propagating (anti)selfdual field configurations of topological charge unity. We also discuss 
kinematic constraints on interacting propagating gauge fields implied by the according spatial coarse-graining, and 
we explain why the screening physics of an SU(2) 
photon is subject to an electric-magnetically dual interpretation. This argument relies on the fact 
that only (anti)calorons of scale parameter $\rho\sim |\phi|^{-1}$ contribute to the coarse-graining required for 
thermal-ground-state emergence at temperature $T$. Thus, use of the effective gauge coupling $e$ in 
the (anti)caloron action is justified, yielding the value $\hbar$ for the latter at almost all temperatures. As a consequence, 
the indeterministic transition of initial to final plane waves caused by an effective, pointlike vertex 
is fundamentally mediated in Euclidean time by a single (anti)caloron being part of the 
thermal ground state. Next, we elucidate how a low-frequency excess of line temperature in the Cosmic Microwave Background 
(CMB) determines the value of the critical temperature of the 
deconfining-preconfining phase transition of an SU(2) Yang-Mills theory postulated 
to describe photon propagation, and we describe how, starting at a redshift of about unity, SU(2) photons 
collectively work 3D temperature depressions into the CMB. 
Upon projection along a line of sight, a given depression influences the present CMB sky in a cosmologically local way, 
possibly explaining the large-angle anomalies confirmed recently by the Planck collaboration. 
Finally, six relativistic polarisations residing in the SU(2) vector modes roughly match the number of degrees of freedom in 
cosmic neutrinos (Planck) which would disqualify the latter as radiation. Indeed, if interpreted as single center-vortex loops 
in confining phases of SU(2) Yang-Mills theories neutrino mass $m_\nu$ solely arises by interactions 
with an environment. Cosmologically, the CMB represents this environment, and thus one would expect that $m_\nu=\xi T$ where $\xi=O(1)$. In this model 
cosmic neutrinos are a small dark-matter contribution, conserved only together with the CMB fluid, influencing 
Baryonic Acoustic Oscillations during CMB decoupling.}
\maketitle
\section{Effective Yang-Mills thermodynamics and SU(2)$_{\mbox{\tiny CMB}}$}

To understand non-Abelian quantum gauge theory requires unconventional thinking because, except for asymptotically large momentum transfer 
\cite{GrossWilczek,Politzer}, the 
fundamental Yang-Mills field $A_\mu$ selfinteracts strongly. In a high-temperature situation a pure SU(N) Yang-Mills theory was argued a 
long time ago to be inaccessible by perturbative techniques 
\cite{Linde}: the so-called soft magnetic sector of the theory is 
screened too weakly for an infrared instability of the small-coupling expansion to be avoided. Namely, starting at order 
six in the perturbative series, kinetic and interaction terms are comparable, contradicting the 
perturbative paradigm that free particle propagation is modified by 
small corrections only. On the other hand, a numerical approach to Yang-Mills 
thermodynamics, which relies on a discretisation of Euclidean spactime on a 4D point 
lattice, suffers from available spatial lattice sizes being too small for capturing the long-range 
correlations of the soft magnetic sector exhaustively \cite{Bielefeld1985}. In contrast to the expectation based on a 
naive substitution of momentum transfer by temperature $T$ in the zero-temperature perturbative running of the 
gauge coupling high-temperature Yang-Mills thermodynamics 
exhibits strong nonperturbative effects \cite{SpatialWilson2008} which, paradoxically (see also \cite{Thierry}), drive the behaviour of propagating 
gauge-field fluctuations in a rapid, power-like way to the Stefan-Boltzmann limit which, however, is subject to a larger number 
of polarisations \cite{Hofmann2005}. Note that the extraction of the nonperturbative lattice $\beta$ function 
is problematic since the perfect lattice action at finite lattice spacing $a$ 
is prohibitively expensive to construct and operate \cite{Hasenfratz and Niedermayer}. 

In \cite{HerbstHofmann2004,Hofmann2005} the deconfining-phase thermodynamics of SU(2) and SU(3) Yang-Mills 
theories was addressed nonperturbatively and analytically. The starting point is 
the construction of unique thermal ground-state estimates in terms of stable, (anti)selfdual, that is, 
nonpropagating gauge-field configurations of topological charge modulus unity: Harrington-Shepard (HS) (anti)calorons \cite{HS1977}. 
A spatial coarse graining over these energy- and pressure-free field configurations in a particular (singular) gauge, 
which is facilitated by the unique computation of the kernel of a 
linear differential operator in terms of an average over space and caloron scale parameter $\rho$ of 
the covariant two-point function of the fundamental (anti)caloron 
field-strength tensor $F_{\mu\nu}$, yields the equation of motion of 
an effective, inert, and adjoint scalar field $\phi$. For later use, it is important to point out that 
the $\rho$ integral is strongly dominated by the upper cutoff \cite{Hofmann2012}. 
The potential of field $\phi$, $V(\phi)=\mbox{tr}\,\Lambda^6/\phi^2$, which represents  
an additively unique entity, is parameterized by a mass 
scale $\Lambda$. The latter emerges by integration of the condition that an Euler-Lagrange type, second-order 
equation and a BPS type, first-order equation simultaneously need to be satisfied by field $\phi$ 
in the coarse-grained equivalent of singular gauge: the winding gauge. As a consequence, $\phi$'s 
modulus is determined as $|\phi|=\sqrt{\frac{\Lambda^3}{2\pi T}}$ in supernatural units $\hbar=c=k_B=1$.  

Apart from potential $V(\phi)$ the complete, Euclidean effective 
action density for the coarse-grained, propagating gauge field $a_\mu$ includes a 
term tr\,$\frac12 G_{\mu\nu}G_{\mu\nu}$, where $G_{\mu\nu}\equiv \partial_\mu a_\nu-\partial_\nu
a_\mu-ie[a_\mu,a_\nu]\equiv G^a_{\mu\nu}\,t_a$ ($e$ the {\sl effective} gauge coupling) and a gauge-invariant 
kinetic term tr\,$(D_\mu\phi)^2$ for field $\phi$. This form of the effective action is uniquely determined 
by the following constraints: (i) coarse-graining over interacting, topologically trivial fluctuations does not lead to a 
change of the form of the fundamental Yang-Mills action \cite{HooftVeltman}, (ii) the effective action needs to be 
gauge invariant, and (iii) field $\phi$ is inert and thus cannot participate in any momentum transfer. While condition (i) forbids the occurrence of higher 
dimensional operators solely constructed from gauge field $a_\mu$, condition (iii) assures 
that, apart from the kinetic term, operators, which mix fields $a_\mu$ and $\phi$, are excluded. From tr\,$(D_\mu\phi)^2$ one reads 
off an adjoint-Higgs-mechanism induced quasiparticle mass term $m^2_a=-2e^2\mbox{tr}\,[\phi,t_a][\phi,t_a]$ associated with direction $a$ of the Lie algebra. For SU(2) and in 
unitary gauge $\phi=2|\phi|\,t_3$, which is reached from winding gauge by an admissible albeit singular 
gauge transformation, one obtains: $m^2\equiv m_1^2=m_2^2=4e^2\frac{\Lambda^3}{2\pi T}$ and $m_3 = 0\,.$ 
Instead of inferring the massiveness of two out of three gauge modes from the action density, one may envisage a Dyson 
series for the propagator of $a_\mu$ subject to an interaction monomial $m^2_a (a_\mu^{a})^2=-2e^2\mbox{tr}\,[\phi,t_a][\phi,t_a](a_\mu^{a})^2$. 
An immediate consequence of field $\phi$ being inert then is that the massive modes $a_\mu^{1,2}$ always propagate with $p^2=m^2$ 
($p$ being the respective four-momentum) because absorption or emission of a finite-momentum mode by 
$a_\mu^{1,2}$, which would facilitate a shift away from this thermal-quasiparticle mass-shell, necessarily transfers 
momentum to field $\phi$. This, however, is excluded by $\phi$'s very definition.  

Demanding that, on the level of noninteracting quasiparticles, the effective theory is thermodynamically 
consistent (Legendre transformations for thermodynamical quantities hold in the effective theory 
as they do in the fundamental theory) yields a first-order evolution equation for $e$ 
as a function of $T$. The respective solution exhibits attractor behaviour, characterised by a plateau $e=\sqrt{8}\pi$ and a 
logarithmically thin pole at the critical temperature $\lambda_c\equiv\frac{2\pi T_c}{\Lambda}=13.87$ of the 
deconfining-preconfining phase transition where monopoles are screened to masslessness and thus start 
to condense \cite{SpatialWilson2008}. Computing radiative 
corrections subject to this $T$ dependence of $e$ yields a tightly controlled loop expansion which is conjectured to 
terminate at a finite irreducible order \cite{Hofmann2006}. The reason for increasing loop orders to imply  
hierarchically decreasing contributions to quantities such as pressure, energy density, and polarisation 
tensors is the explosion of the number of independent kinematic constraints with respect to 
the number of integration variables. Kinematic constraints, which are simple to formulate in physical, 
Coulomb-unitary gauge (total Cartan and off-Cartan gauge fixing, respectively), arise because the momentum 
transfer through an effective four-vertex is contrained to be $|\phi|^2$ in each Mandelstam variable 
$s$, $t$, and $u$ and because the massless mode's four momentum $p$ needs to 
satisfy $|p^2|\le |\phi|^2$. Note that scattering, mediated by a four-vertex in the effective theory, is computed as a 
coherent average over the contributions from each channel \cite{KrasowskiHofmann2013}. 

Let us now review the role of coupling $e$ in the effective theory \cite{Hofmann2012}. 
Working in units where $c=k_B=1$ but $\hbar$ is re-instated as an action, the (dimensionless) exponent 
\eqb
\label{expweight}
-\frac{\int_0^{\beta} d\tau d^3x\,{\cal L}^\prime_{\mbox{\tiny eff}}[a_\mu]}{\hbar}\,, 
\eqe
in the weight belonging to fluctuating fields in the effective partition function\footnote{Since $\phi$ is inert the factor, 
whose exponent is the potential-part of the effective action, can be pulled out of the partition 
function formulated in terms of effective fields.} can, in unitary gauge, be re-cast as 
\eqb
\label{expweightdiml}
-\mbox{tr}\,\int_0^{\beta} d\tau d^3x\,\Big(\frac12(\partial_\mu \tilde{a}_\nu-\partial_\nu
\tilde{a}_\mu-ie\sqrt{\hbar}[\tilde{a}_\mu,\tilde{a}_\nu])^2-e^2\hbar[\tilde{a}_\mu,\tilde{\phi}]^2\Big)\,, \ \ \ \ \ \ \ (\beta\equiv\frac{1}{T})\,,
\eqe
if we define $\tilde{a}_\mu\equiv a_\mu/\sqrt{\hbar}$ and $\tilde{\phi}\equiv \phi/\sqrt{\hbar}$ and assume these fields not to 
depend on $\hbar$ \cite{Brodsky2011}. Because of terms  
$\propto \hbar^0$ in (\ref{expweightdiml}) the unit of $\tilde{a}_\mu$ is 
length$^{-1}$ which is also true of $\tilde{\phi}$. Thus coupling $e$ has units of $1/\sqrt{\hbar}$, and we have
\eqb
\label{plateaucoupl}
e=\frac{\sqrt{8}\pi}{\sqrt{\hbar}}
\eqe
almost everywhere in the deconfining phase. Only (anti)calorons of scale parameter 
$\rho\sim |\phi|^{-1}$ contribute to the emergence of field $\phi$ and thus to the thermal ground 
state as a whole. In other words, only those (anti)calorons, which takes place just below the level of resolution in the 
effective theory, actually support the description of deconfining Yang-Mills thermodynamics in terms 
of fields $\phi$ and $a_\mu$. Therefore, it is justified to use the effective coupling $e$ when computing 
the action of a single, {\sl relevant} (anti)caloron as  $S_{\tiny\mbox{(A)C}}=\frac{8\pi^2}{e^2}$. But 
Eq.\,(\ref{plateaucoupl}) then implies that $S_{\tiny\mbox{(A)C}}=\hbar$ which suggests that the indeterminism of plane-wave scattering by isolated 
effective vertices occurs because in and out states are interpolated in {\sl imaginary time} by a fundamental, topologically 
nontrivial field configuration of a definite action.       

In \cite{Hofmann2005} it was assumed that the critical temperature $T_c$ for the deconfining-preconfining phase transition of a 
particular SU(2) Yang-Mills theory associated with 
thermal photon propagation coincides with the present baseline temperature $T_0=2.725\,$K of the Cosmic Microwave Background (CMB). 
This speculation is supported by the balloon borne NASA mission Arcade 2 \cite{Arcade2} 
and earlier terrestial radio-frequency CMB radiance measurements, see references in \cite{Arcade2}. 
Their results indicate a highly significant increase of CMB line 
temperature for decreasing frequencies below 3\,GHz which was interpreted in \cite{Hofmann2009} as a mild redistribution of 
the CMB Rayleigh-Jeans spectrum due to an onset of photon Meissner massiveness and an accompanying 
evanescence of low frequency modes. For the SU(2) photon to acquire a small Meissner mass ($m_\gamma\sim 100\,$MHz), as suggested by the data,  
one requires $T_c$ to be slightly below $T_0$. For all practical purposes, however, one can set 
$T_c=T_0$, and we have introduced the names SU(2)$_{\mbox{\tiny CMB}}$ for the physical 
Yang-Mills theory of this scale and $V^\pm$ for the vector modes of this model whose common 
mass vanishes with $T^{-1/2}$ with increasing temperature $T$.    

This paper is organised as follows: The next section gives arguments on why the SU(2)$_{\mbox{\tiny CMB}}$ photon 
and the topological defects of SU(2)$_{\mbox{\tiny CMB}}$ (in the deconfining phase: unresolved, screened monopoles; 
in the preconfining phase: condensed massless monopoles) enjoy an electric-magnetically dual interpretation. Also, results on the longitudinal and transverse 
parts of the polarisation tensor of the massless mode are reviewed. 
In Sec.\,\ref{4} the afore mentioned low-frequency anomaly of CMB line temperature is described in 
more detail. Sec.\,\ref{5} exposes the major CMB anomalies at large angles as recently confirmed by the results of the 
Planck collaboration. A scenario on how SU(2)$_{\mbox{\tiny CMB}}$ can possibly explain these anomalies is 
discussed in Sec.\,\ref{6}. Finally, Sec.\,\ref{7} is devoted to the apparent problem of SU(2)$_{\mbox{\tiny CMB}}$ yieding too many relativistic 
degrees of freedom appear at big-bang nucleosynthesis and CMB decoupling: Already at moderate redshifts the 
SU(2)$_{\mbox{\tiny CMB}}$ vector modes $V^\pm$ nearly match the number of massless neutrino degrees 
of freedom as extracted from the Planck data. This leads to the proposition that the latter should be replaced by the former. Cosmic 
neutrinos, in turn, are argued to behave like (luke-)warm dark matter: Interpreted as single center-vortex loops in the confining phases 
of additional SU(2) Yang-Mills theories, their mass emerges by interaction with the CMB.             
     
\section{Electric-magnetically dual interpretation in SU(2)$_{\mbox{\tiny CMB}}$\label{2}}

If photon progation is, indeed, due to an SU(2) rather than a U(1) gauge principle 
then the observation that $e=\frac{\sqrt{8}\pi}{\sqrt{\hbar}}$ has implications. Namely, the question then 
arises how the unit of electric charge $Q$ is seen by the SU(2)$_{\mbox{\tiny CMB}}$ photon as 
compared to a conventional U(1) photon. 
In units of $c=k_B=1$ and with $\hbar$ re-instated as an action, this is addressed by recalling that the 
QED fine-structure constant $\alpha$, a unitless quantity, 
is given as 
\begin{equation}
  \label{massessu2} 
\alpha=\frac{Q^2}{4\pi\hbar}\,.
\end{equation}  
For $\alpha$ to be unitless (in any system of units $\alpha$ is dimensionless) 
$Q$ should be proportional to $\sqrt{\hbar}$ which is the case if $Q\propto 1/e$. Since this is true of {\sl magnetic} 
charge w.r.t. to the Cartan subgroup U(1) of SU(2) (the charge of a monopole inside an (anti)caloron of 
extent $\rho\sim |\phi|^{-1}$ is $\frac{4\pi}{e}$) we conclude that an electric field of the real world actually is a magnetic field 
in U(1)$\in$SU(2)$_{\mbox{\tiny CMB}}$ and vice versa. That is, an electric-magnetically dual interpretation 
of electromagnetism based on SU(2)$_{\mbox{\tiny CMB}}$ is in order\footnote{The Bianchi 
identities of the conventional theory are not violated since 
the vector modes $V^\pm$ of SU(2)$_{\mbox{\tiny CMB}}$, which are of magnetic charge in the real world, practically do not couple to 
the photon at electric field strengths prevailing in atoms and molecules and also for macroscopic electric 
field strengths, see discussion in \cite{FalquezHofmannBaumbach2011}.}. Thus SU(2)$_{\mbox{\tiny CMB}}$ predicts the existence of 
electric rather than magnetic monopoles. These monopoles, however, are never resolved in the effective theory: 
they collectively contribute to the alteration of conventional, thermal photon propagation.

\section{Polarisation tensor of the thermal photon\label{3}}

Here we review briefly how (anti)screening effects due to interaction with the vector modes $V^\pm$ influence the propagation properties 
of the photon $\gamma$ in SU(2)$_{\mbox{\tiny CMB}}$. There are two invariants $G(p_0,\vec{p},T)$ and $F(p_0,\vec{p},T)$ 
which modify the respective dispersion laws for tranverse and longitudinal\footnote{On tree level, 
there is no longitudinal polarisation, but magnetic charge density waves are generated radiatively in the Yang-Mills plasma.} 
photon propagation of four momentum $(p_0,\vec{p})$ in a thermal plasma 
containing $\gamma$ and $V^\pm$. Collectively, interaction of $\gamma$ with $V^\pm$ in the effective 
theory summarises screening and antiscreening effects the photon experiences due to the presence of {\sl unresolved} electric monopoles 
induced into the thermal ground-state estimate by effective radiative corrections \cite{SpatialWilson2008}. On a fundamental level, 
induction of monopole constituents is a consequence of transient holonomy shifts in (anti)calorons \cite{Nahm,LeeLu,VanBaal,Diakonov} 
due to their interaction with the propagating sector. Depending on the acquired holonomy, induced monopole-antimonopole pairs 
either attract one another (small holonomy), leading to annihilation, or, rarely, their repulsion (large holonomy) causes the 
(anti)caloron to dissociate into these constituents which then are subject to screening by other (anti)monopoles. We have argued and shown in 
\cite{Hofmann2006,SchwarzHofmannGiacosa2007,KavianiHofmann2007} that due to a large hierarchy in radiative corrections with increasing 
loop order it is sufficient for practical purposes to compute $\gamma$'s polarisation tensor $\Pi_{\mu\nu}$ 
on the one-loop level.   

\begin{figure}
  \centering
  \includegraphics*[width=\textwidth]{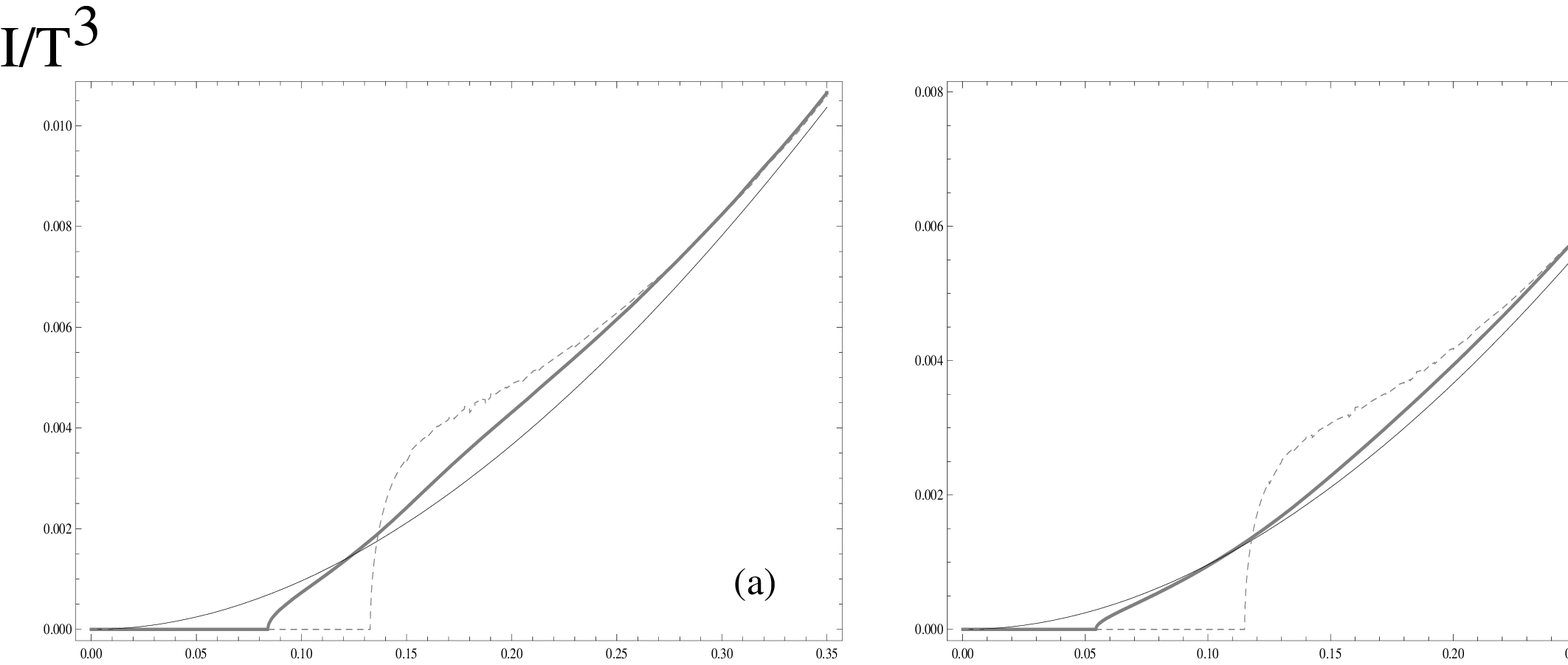}
  \caption{\protect{\label{Fig-2}} Plots of the results for the
    dimensionless spectral energy density $\frac{I}{T^3}$ obtained in
    the one-loop selfconsistent calculation (solid grey curves) and
    with the approximation $p^2=0$ (dashed grey curves) for (a)
    $T=3\,T_c$ and (b) $T=4\,T_c$. Here $Y\equiv\frac{p_0}{T}$.
    Solid black curves depict the conventional Rayleigh-Jeans spectrum. Note
    the much faster approach with rising temperature to the low-frequency Planck
    spectrum in the selfconsistent one-loop result in comparison with the
    approximate calculation.}
\end{figure}

In unitary-Coulomb gauge kinematic constraints allow only diagram B in Fig.\,\ref{Fig-1} to contribute to $\Pi_{\mu\nu}$. 
For screening function $G$ the following gap equation arises \cite{SchwarzHofmannGiacosa2007,Ludescher2008}:
\begin{figure}
\centering
\includegraphics*[width=\textwidth]{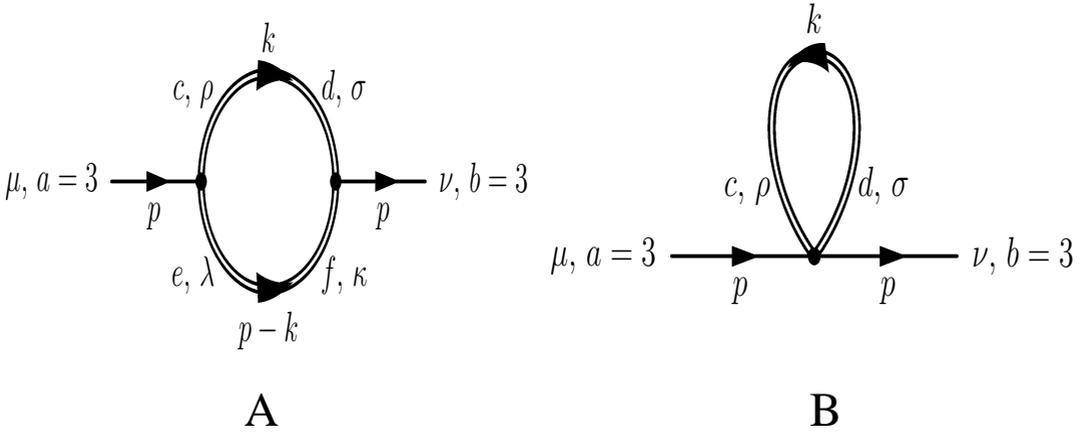}
\caption{The two diagrams potentially contributing to the one-loop polarisation tensor of $\gamma$. Double lines associate  
with $V^\pm$ propagation, a single line with $\gamma$.}
\label{Fig-1}     
\end{figure}    
\eab
\label{dimlPixx}
\frac{G}{T^2}&=&\frac{\Pi^{11}}{T^2}=\frac{\Pi^{22}}{T^2}=\nonumber\\ 
&=&\frac{e^2}{\pi\lambda^3}\,\int
d^3y\,\left(-2+\frac{y_1^2}{4e^2}\right)\,
\frac{n_B\left(2\pi\lambda^{-3/2}\sqrt{\vec{y}^2+4e^2}\right)}{\sqrt{\vec{y}^2+4e^2}}\,,
\eae
where the integration over dimensionless variables on the right-hand side is subject to the constraint
\eqb
\label{constcyl}
\left|\frac{G}{T^2}\frac{\lambda^{3}}{(2\pi)^2}+\frac{\lambda^{3/2}}{\pi}\left(\pm\sqrt{X^2+
\frac{G}{T^2}}\sqrt{\rho^2+\xi^2+4e^2}-X\xi\right)+4e^2\right|\le 1\,.
\eqe
In (\ref{constcyl}) cylindrical coordinates $y_1\equiv\rho\,\cos\varphi$, $y_2\equiv\rho\,\sin\varphi$\,, and $y_3=\xi$ 
were introduced, and we define $X\equiv\frac{|\vec{p}|}{T}$. 
Similar expressions hold for $F$'s gap equation \cite{FalquezHofmannBaumbach2011}. The modification of transverse photon dispersion introduced 
by $G$ leads to an anomaly in the blackbody spectrum of energy density which is indicated in Fig.\,\ref{Fig-2} \cite{SchwarzHofmannGiacosa2007,Ludescher2008}.  
Longitudinal propagation occurs in various low-momentum branches. In \cite{FalquezBaumbachHofmann2012} 
we have argued that these incoherent magnetic-field modes can acquire large correlation lengths if 
temperature $T$ exhibits an (adiabatically slow) dependence on space. In Sec.\,\ref{6} this will be shown to 
happen on cosmological distances associated with a reshift of order unity by induction through the blackbody 
anomaly in transverse modes. So the longitudinal $\gamma$ sector could provide for emergent cosmic magnetic fields by virtue of 
CMB temperature inhomogeneities arising from anomalous thermal photon propagation.          

\section{CMB low-frequency anomaly and SU(2)$_{\mbox{\tiny CMB}}$\label{4}}

To make contact with CMB observation it is necessary to fix the Yang-Mills scale or, which is the same, $T_c$ 
of SU(2)$_{\mbox{\tiny CMB}}$ by observation. To this end, we discuss low-frequency observations 
of the CMB line temperature. 

Arcade 2 is a balloon borne instrument to observe the CMB at very low 
frequencies (bands at $\nu=3,8, 10, 30,$ and $90\,$GHz) by alternating observation of a calibrator black body emitter and 
the CMB sky. After subtraction of known and well modelled foreground radiance 
in the sky signal, an adjustment of the calibrator's temperature such 
that the two signals null each other yields the temperature of the calibrator as a 
function of frequency (CMB line temperature). The interesting feature in this data was the 
detection of a clear excess (statistically significant at the level of five standard deviations) 
of the CMB line temperature $T(\nu)$ at 3 and 8\,GHz. This is expressed by a power-law model \cite{Arcade2} 
\eqb
\label{lineT}
T(\nu)=T_0+T_R\,\left(\frac{\nu}{\nu_0}\right)^\beta\,.
\eqe
subject to the following parameter values: $T_0=2.725\,$K (within errors FIRAS' CMB baseline 
temperature \cite{FIRAS1994}) obtained by a fit to the CMB spectrum at frequencies from 60\,GHz to 600\,GHz), 
$\nu_0=310\,$MHz, $T_R=24.1\pm 2.1\,$K, and $\beta=-2.599\pm 0.036$. The claim by the collaboration 
that this significant low-frequency deviation from a perfect black-body spectrum  ($T(\nu)\equiv \mbox{const}$) 
is not an artefact of galactic foreground subtraction, unlikely is related to an average effect of 
distant point sources, and that these results confirm the trend seen in 
earlier radio-frequency data \cite{Roger1999,Maeda1999,Haslam1981,Reich1986} 
appears to rest on solid grounds. In \cite{Hofmann2009} this 
observational situation was taken as motivation for an unconventional explanation of Eq.\,(\ref{lineT}) which, 
at the same time, determines the value of $T_c$ for SU(2)$_{\mbox{\tiny CMB}}$. 

Roughly speaking, the rise in line temperature can be explained if $\gamma$ modes become evanescent at low 
frequencies due to the onset of the Meissner effect (electric monopoles start to condense into a new ground state at $T_0$). 
This leads to a re-arrangement of spectral blackbody power at low frequencies: the power of evanescent modes is maximal at 
$\nu=0$ and matches that of propagating photons at a transition frequency which is comparable to the (effective and feeble
\footnote{In \cite{Hofmann2009} it was assumed 
that monopole condensation at $T_0$ is not sufficiently strong to lead to an extra photon polarisation.}) photon mass 
of about 100\,MHz. In SU(2)$_{\mbox{\tiny CMB}}$ this scenario takes place if $T_c$ is slightly below $T_0$, and, for all practical purposes, 
one may set $T_c=T_0$.            

\section{CMB large-angle anomalies: WMAP and Planck\label{5}}

Recently released data by the Planck collaboration confirm and strengthen the occurrence of 
large-angle anomalies in the CMB map of temperature fluctuations \cite{AdeI} which were first 
noticed based on WMAP data \cite{Tegmark,Oliveira-Costa,Copi,Vielva}. Here are some of them:  
\begin{itemize}
\item 
Large-angle suppression of $TT(\theta)$, defined as the product of temperature in two pixels, 
which are separated by angle $\theta$, 
averaged over all possible pixel pairs. On angular scales $\theta > 60^\circ$ 
and most notably on the northern ecliptic hemisphere (equatorial plane = plane of the solar system), 
$TT(\theta)$ is significantly smaller than predicted by the Cosmological Standard Model (CSM).
\item 
Low variance of temperature fluctuations in the northern compared to the southern 
ecliptic hemisphere. The variance in the latter essentially agrees with CSM simulations. 
\item 
Power asymmetry. There is a plane, again close to the ecliptic, for which the 
difference between the average amplitude of temperature fluctuations in the northern and southern hemispheres 
reaches a maximum. 
\item 
The CMB ‘cold spot’ — a region of anomalously low temperature. 
The cold spot is detected by an estimation of the kurtosis of the distribution of 
temperature values in its surroundings, indicating a local violation of Gaussianity. 
\item 
Significant alignment of the CMB quadrupole with the octupole.  
\item 
Hemispherical antisymmetry. Temperature fluctuations in pixels that are 
related by mirror-reflection about a plane, whose normal is misaligned with the cold-spot direction by about 42$^\circ$, 
tend to be the negative of one another. 
\end{itemize}
A dipolar modulation model \cite{Gordon}, which assumes a statistically isotropic CMB sky to be multiplicatively 
influenced by a vector $\vec{p}$, mimicks the actual CMB sky realistically \cite{AdeI}.      

\section{Dynamical breaking of the CMB's statistical isotropy by SU(2)$_{\mbox{\tiny CMB}}$ \label{6}}

\begin{figure}
  \centering
  \includegraphics[width=0.65\textwidth]{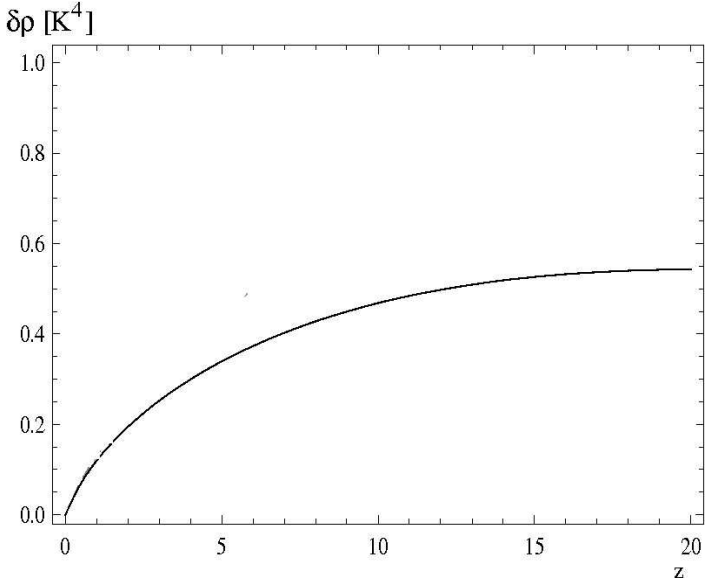}
  \caption{\protect{\label{Fig-3}} The difference $\delta\rho$ in energy densities of 
the $\gamma$ part in SU(2)$_{\mbox{\tiny CMB}}$ and of thermal photons in the conventional U(1) theory for photon propagation as a function of 
$z\equiv T/T_0-1$.}
\end{figure}

To discuss how SU(2)$_{\mbox{\tiny CMB}}$ could explain the CMB 
large-angle anomalies it is important to consider the difference $\delta\rho$ 
in energy densities, given by the respective spectral integrals, of the $\gamma$ part in SU(2)$_{\mbox{\tiny CMB}}$ and of thermal photons in the 
conventional U(1) theory for photon propagation, see Fig.\ref{Fig-2}, as a function of $T$. Fig.\,\ref{Fig-3} depicts $\delta\rho(T)$. Note that function 
$\delta\rho$ exhibits {\sl positive} slope which is maximal at $T_c$.
In propagating the CMB from decoupling at redshift $z\sim 1100$ to today ($z=0$) 
its thermal photons should stream freely according to conventional theory modulo effects induced by 
nonlinear growth of matter structure (gravitational lensing, integrated Sachs-Wolfe) and 
re-ionised interstellar hydrogen/stellar plasmas (Sunyaev-Zeldovich). Disregarding these small effects, observed
CMB temperature fluctuations $\delta T$ should be determined (i) 
by initial conditions, transferred from those of the matter sector through linear theory, and (ii) 
baryonic acoustic oscillations (BAO), influencing the CMB during decoupling. Note that angular fluctuations 
$\delta T$ are projections along lines of sights through 3D temperature distributions 
$\Delta T (z,\vec{x})$ for redshifts $z$. 

If distribution $\Delta T$ is discretised on physical volume elements $\Delta V$, proportional to the third power of the FRW scale factor $a$,     
then a thermodynamical a priori probability density $P_z(\Delta T)$ for $\Delta T$ to occur 
at redshift $z$ within volume element $\Delta V$ can be assigned in the sense of a canonical ensemble (units are: $k_B=\hbar=c=1$)
\eqb
\label{aprioriProb}
P_z(\Delta T)=\frac{\exp\left(-\frac{\rho(\bar{T}+\Delta T)\Delta V}{\bar{T}}\right)}{\int_{T_0}^\infty dT\,
\exp\left(-\frac{\rho(T)\Delta V}{\bar{T}}\right)}\,,
\eqe 
where now redshift $z$ relates to the average CMB temperature $\bar{T}$ as 
$z=\frac{\bar{T}}{T_0}-1$. Note that since the Silk cutoff for BAO translates today into a spatial scale equal to a typical 
galaxy diameter and because the energy density of today's CMB is $\rho_{\tiny\mbox{U(1)}}\sim \frac{\pi^2}{15}\,T_0^4=\frac{\pi^2}{15} (11.9\,\mbox{cm}^{-1})^4$ the 
magnitude of the exponents in Eq.\,(\ref{aprioriProb}) is very large such that two very small numbers appear in the numerator and denominator 
defining $P_z(\Delta T)$. The factor $F_z$ to bias temperature fluctuations $\Delta T$ for SU(2) effects in 
cooling down the CMB by cosmological redshift should be 
the probability for $\Delta T$ to occur in volume element $\Delta V$ in SU(2)$_{\mbox{\tiny CMB}}$ normalized to the according U(1) probability
\footnote{In CMB simulations fluctuation $\Delta T$ is determined nonthermally in terms of Gaussian initial conditions, but $P_z(\Delta T)dT$ states 
how likely an apparent temperature parameter in a region represented by volume element $\Delta V$ would assume values in the range from 
$\bar{T}+\Delta T-\frac12 dT$ to $\bar{T}+\Delta T+\frac12 dT$ given an overall heat reservoir of physical, mean temperature $\bar{T}$. So long as there is no 
such physical reservoir, which is true of CMB photons after their last scattering if described by conventional U(1) gauge theory but also for
 SU(2)$_{\mbox{\tiny CMB}}$ photons if $\bar{T}\gg T_0$, $P_z(\Delta T)$ is an invalid concept. This is accomodated by the cancelation 
to unity between SU(2)$_{\mbox{\tiny CMB}}$ and U(1) in the definition of $F_z(\Delta T)$ in Eq.\,(\ref{SU(2)vsU(1)}). 
However, for $\bar{T}$ larger but comparable to 
$T_0$ the thermal ground state of SU(2)$_{\mbox{\tiny CMB}}$ and the thermal ensemble of $V^\pm$ modes 
couple to $\gamma$ and thus commonly act as a heat reservoir. In the ratio $F_z(\Delta T)$ of $P_z(\Delta T)$ for SU(2)$_{\mbox{\tiny CMB}}$ and U(1) 
the U(1) part in SU(2)$_{\mbox{\tiny CMB}}$ still cancels. This part is due to non-Rayleigh-Jeans frequencies, which do not 
lend themselves \cite{Ludescher2008,FalquezHofmannBaumbach2011} to the concept of Eq.\,(\ref{aprioriProb}).}:
\eqb
\label{SU(2)vsU(1)}
F_z(\Delta T)=\frac{P_{z,\tiny\mbox{SU(2)}}(\Delta T)}{P_{z,\tiny\mbox{U(1)}}(\Delta T)}\propto \exp\left(-\frac{\delta\rho(\bar{T}+\Delta T)
\Delta V}{\bar{T}}\right)\,.
\eqe
Since $\delta\rho$ has positive slope, which vanishes, however, in a powerlike way for $\bar{T}\to\infty$ (see Fig.\,(\ref{3})), {\sl negative} fluctuations 
$\Delta T$ are for $0\le z\le 1$ significantly favoured over positive ones ($F_z(-|\Delta T|)>F_z(|\Delta T|)$) by prescribing  
\eqb
\label{biasing}
\Delta T\to F_z(\Delta T)\Delta T\,.
\eqe
If the propagation of the CMB to $z=0$ is performed in Fourier space then the multiplication of (\ref{biasing}) turns into 
a convolution operation.    

What are the implications? If by primordial chance $\Delta T_{\vec x}$ is negative within a given 
volume element $\Delta V_{\vec x}$ ($\vec{x}$ denoting a co-moving location at the center of $\Delta V_{\vec x}$) 
at a redshift $z\sim 1$ then the biasing of (\ref{biasing}) strengthens this 
trend. That is, a local dip in temperature becomes deeper and, in going to lower redshift, broader due to the free-streaming part of the 
CMB photon spectrum cooling off the region contained in the forward light cone of $\vec x$. This leads to the emergence of 
cold regions of extent $z\sim 1$, their centers representing the locations 
of exceptionally negative, primordial seed fluctuations $\Delta T_{\vec x}$. Observation of the CMB from a present position at the slope of one  
of these cold regions or temperature depressions faces a preferred direction: the local gradient $\vec{p}$ of such a temperature depression. Statistical isotropy 
thus is broken locally by SU(2)$_{\mbox{\tiny CMB}}$ thermodynamics. It is clear that performing projections 
along and counter to the direction of $\vec{p}$ yields maximally asymmetric results: In the former case more unbiased fluctuations contribute, 
because the distance to the edge of the depression is small, and a variance asymmetry is expected as a consequence. 
Also, a dynamical contribution to the CMB dipole arises from the difference of these extremal projections while a 
cold spot emerges from their average. Because $\vec{p}$ emerges for $z<1$ it is clear that primordial fluctuations on large 
angular scales are most affected by the build-up of an observable temperature depression: this process suppresses the temperature-temperature correlation 
function for large angles, and it aligns low multipoles.

\section{Cosmic neutrinos\label{7}}

Let us finally address an interesting match of the non-photonic number of relativistic degrees of 
freedom extracted from the high-$l$ part of the angular power spectrum associated with the 
Planck data: \cite{AdeII} reports a number $N_{\tiny\mbox{eff}}$ of neutrino flavours as 
$N_{\tiny\mbox{eff}}=3.30\pm0.27$. Interestingly, this implies a number of about six relativistic degrees 
of freedom at CMB decoupling which is a near match to those represented by $V^\pm$ of SU(2)$_{\mbox{\tiny CMB}}$: three relativistic 
polarisations in each of the two vector species. Note that already for $\bar{T}$ not much larger 
than $T_0=T_c$, the ratio $R$ of the (common) mass of $V^\pm$ to ${\bar T}$ is given as  
\eqb
\label{Vpmmass}
R=\frac{m_{V^\pm}}{\bar{T}}=2e\frac{|\phi|}{\bar{T}}=\sqrt{108}\pi^2\lambda_c^{-3/2}\left(\frac{\lambda}{\lambda_c}\right)^{-3/2}\,.
\eqe
For $\frac{\lambda}{\lambda_c}=2,10,1100$ one has $R=0.702,0.063,5.4\times 10^{-5}$, respectively, and 
thus $V^\pm$ describe relativistic degrees of freedom already at moderate redshifts and certainly so 
during CMB decoupling. But if during CMB decoupling the vector modes $V^\pm$ of SU(2)$_{\mbox{\tiny CMB}}$ 
actually play the role conventionally attributed to neutrinos 
then how do the latter vanish from the spectrum of relativistic degrees of freedom? 
In \cite{Planckscaleaxion,CVL2008} we have proposed that neutrinos could be single, that is, non-selfintersecting 
center-vortex loops\footnote{A closed chain of nearly massless monopoles and antimonopoles, 
which, due to an electric-magnetically dual interpretation, carry feeble electric charges, locally move along and opposite to the 
tangential direction of the loop, and generate one unit of magnetic center flux.} 
(zero selfintersection number, $N=0$) in the confining phases of pure SU(2) gauge theories whose Yang-Mills scales 
match the masses of charged leptons. Due to the absence of an explicit mass scale in the sector with $N=0$ and void of interactions with 
an environment a single vortex loop would shrink to a round point \cite{CVL2008} of zero mass and vanishing electric/magnetic multipoles. 
However, cosmic neutrinos do interact with an environment, the CMB, over long time scales. Already at moderate temperatures, 
the only CMB mass scale is $\bar{T}$, and thus we expect that neutrino masses are given as $m_\nu=\xi\bar{T}$ where $\xi$ is a dimensionless 
factor of order unity macroscopically parametrising the effect of the CMB on the mass of a center-vortex loop. 

\begin{figure}
  \centering
  \includegraphics[width=0.65\textwidth]{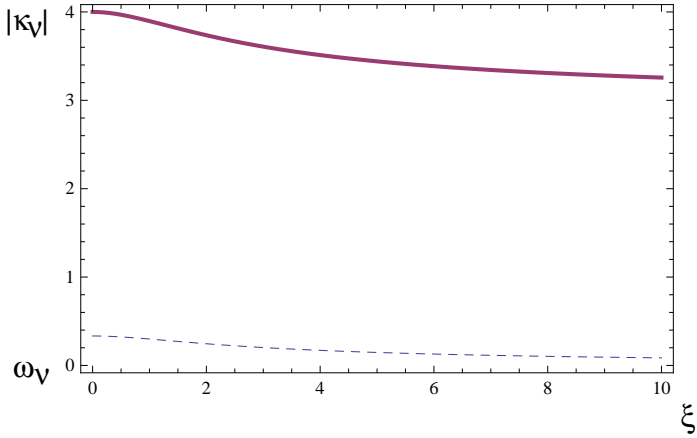}
  \caption{\protect{\label{Fig-4}} The modulus of $\kappa_\nu$ in $\rho(a)=\rho_0 a^{\kappa_\nu}$ (solid line), where $\rho$, $a$ denote 
energy density, scale factor, respectively ($a_0\equiv 1$), and $\omega_\nu$ in $p=\omega_\nu\rho$ (dashed line), where $p$ denotes pressure, as 
functions of $\xi$ for cosmic neutrinos if the latter are assumed to be single center-vortex loops of pure SU(2) gauge theories 
in their confining phases. In this model, neutrinos would exhibit a CMB induced mass $m_\nu=\xi\bar{T}$.}
\end{figure}

Is this compatible with neutrino oscillations physics? The strongest bound of $8\times 10^{-3}\,$eV on 
neutrino mass scales comes from solar neutrino oscillations \cite{Maltoni}. Since $T_0=2.35\times 10^{-4}\,$eV 
this bound leads to the constraint $\xi\le 30$. But already for values of $\xi$ moderately larger than unity neutrinos 
would behave like matter rather than radiation, see Fig.\,\ref{Fig-4}. Note that in this scenario 
CMB and neutrino fluids are no longer separately conserved.  
In CMB simulations $\xi$ should be included as a new fit parameter, and we expect a small, (luke-)warm contribution to dark 
matter to be mimicked by the neutrino sector during decoupling which should have an influence on BAO. 
Since the here-discussed scenario assumes that a thermal neutrino mass emerges by feeble interactions with the CMB the comparatively very 
local freeze-out physics at the onset of big-bang nucleosynthesis, where relative neutron and proton number densities are fixed 
by the equality of their conversion rate and the rate of 
expansion of the Universe, is not influenced: thermal masses of cosmic 
neutrino are induced after the fact, that is, only after their decoupling from matter.    

\section{Summary and conclusion\label{8}}

In this talk we have briefly reviewed the nonperturbative physics leading to the emergence of a 
thermal ground state in the deconfining phase of an SU(2) Yang-Mills theory, and we have discussed 
why the gauge symmetry breaking SU(2)$\to$ U(1), dynamically induced by this ground state -- 
collectively representing calorons and anticalorons of topological charge modulus unity --, is commensurate with 
the postulate that photon propagation is fundamentally described by an SU(2) rather 
than a U(1) gauge principle, see also \cite{RalfNP}. In this context, explanations of a low-frequency anomaly in CMB line temperature \cite{Arcade2} and 
recently confirmed \cite{AdeI} large-angle anomalies in the temperature fluctuations of the CMB sky 
are proposed which entail a re-interpretation of the number of relativistic degrees of freedom 
during CMB decoupling.

\section*{Acknowledgments}
We would like to acknowledge useful conversations with James Annis and Markus Schwarz.

\end{document}